\documentclass [twocolumn] {aastex62}
\usepackage{graphicx,color}   
\newsavebox\mysavebox
\usepackage{color}            
\usepackage{natbib}
\usepackage{hyperref} 
\hypersetup{linkcolor=blue,citecolor=blue,filecolor=cyan,urlcolor=blue}
\usepackage{amsmath}
\usepackage{lineno}
\usepackage{pstricks}

\newcommand{\logg}{log $\it{g}$}
\newcommand{\teff}{$T_{eff}$}
\graphicspath{{./}{figures/}}
\received{\today}
\revised{}
\accepted{}
\submitjournal{AJ}
\shorttitle{carbon abundance}
\shortauthors{Athira et al}

\begin{document}

\title{Carbon abundance of stars in the LAMOST-Kepler field}

\author[0000-0001-6093-5455 ]{Athira Unni }
\affil{Indian Institute of Astrophysics,Koramangala 2nd Block, Bangalore 560034, India}
\affil{Pondicherry University, R.V. Nagar, Kalapet, 605014, Puducherry, India}
\author[0000-0002-0554-1151 ]{Mayank Narang}
\affiliation{Department of Astronomy and Astrophysics, Tata Institute of Fundamental Research, Homi Bhabha Road, Colaba, Mumbai 400005,India}

\author[0000-0003-0891-8994 ]{Thirupathi Sivarani}
\affiliation{Indian Institute of Astrophysics,Koramangala 2nd Block, Bangalore 560034, India}

\author[0000-0002-3530-304X]{Manoj Puravankara}
\affiliation{Department of Astronomy and Astrophysics, Tata Institute of Fundamental Research, Homi Bhabha Road, Colaba, Mumbai 400005,India}

\author[0000-0003-0799-969X ]{Ravinder K Banyal }
\affiliation{Indian Institute of Astrophysics,Koramangala 2nd Block, Bangalore 560034, India}

\author[0000-0002-9967-0391]{Arun Surya}
\affiliation{Department of Astronomy and Astrophysics, Tata Institute of Fundamental Research, Homi Bhabha Road, Colaba, Mumbai 400005,India}

\author[0000-0003-0003-4561]{S.P. Rajaguru}
\affiliation{Indian Institute of Astrophysics,Koramangala 2nd Block, Bangalore 560034, India}

\author[0000-0003-1371-8890]{C.Swastik}
\affiliation{Indian Institute of Astrophysics,Koramangala 2nd Block, Bangalore 560034, India}
\affil{Pondicherry University, R.V. Nagar, Kalapet, 605014, Puducherry, India}

\begin{abstract}
The correlation between host star iron abundance and the exoplanet occurrence rate is well-established and arrived at in several studies. Similar correlations may be present for the most abundant elements, such as carbon and oxygen, which also control the dust chemistry of the protoplanetary disk. In this paper, using a large number of stars in the Kepler field observed by the LAMOST survey, it has been possible to estimate the planet occurrence rate with respect to the host star carbon abundance. Carbon abundances are derived using synthetic spectra fit of the CH G-band region in the LAMOST spectra.
The carbon abundance trend with metallicity is consistent with the previous studies and follows the Galactic chemical evolution (GCE).
Similar to [Fe/H], we find that the [C/H] values are higher among giant planet hosts. The trend between [C/Fe] and [Fe/H] in planet hosts and single stars is similar; however, there is a preference for giant planets around host stars with a sub-solar [C/Fe] ratio and higher [Fe/H]. Higher metallicity and sub-solar [C/Fe] values are found among younger stars as a result of GCE. Hence, based on the current sample, it is difficult to interpret the results as a consequence of GCE or due to planet formation. 
\end{abstract}

\section{Introduction} \label{sec:intro}

Planets and their host stars are formed together from the same molecular cloud. Naturally, the planet's chemical composition is expected to correlate with the host star. Hence, studies of the host star's chemical abundances could constrain the planet's bulk abundance and the planet formation process.
 Host star metallicity and giant planet connection was first observed by  \citet{planet_metalicity_gonzalez, gonzalez1998} and confirmed by \citet{Santos2001, Santos2004b} with a larger sample, and these authors also showed that the frequency of giant planet hosts increases steeply above solar metallicity. This rapid rise in giant planet (R$_{p}>4$R$_{\oplus}$) occurrence of 3\% at solar metallicities and up to 25\% at [Fe/H]=0.3 was again shown by \citet{ planet_metalicity}. \citet{planet_metalicity1} observed giant planet-metallicity correlation in a wide range of stellar masses, and the occurrence increased from 3\% in M dwarfs to 14\% in A dwarfs at solar metallicity. Although the metallicity trend was absent for stars that host smaller planets, a large spread in metallicities is observed among them\citep{Sousa2008,Neves2009} and the low-mass planet-bearing stars at low metallicity were found to be rich in $\alpha$ elements \citep{Adibekyan2012a,Adibekyan2012c,Adibekyan2012b,fgk2}. \citet{Adibekyan2012a} suggested terrestrial planets could form early in the Galaxy among the thick disk stars due to their enhanced $\alpha$ abundances. A recent study by \citet{swastik_2022} showed that [$\alpha$/Fe] ratio shows a negative trend with respect to planetary mass, indicating possible conditions for the formation of low-mass planets before Jupiter-like planets. 
The host star mass metallicity trend was also found to reverse for planet masses higher than M $>$ 4M$_{J}$ \citep{mayank}. Directly imaged planets also showed a large scatter in the metallicity among super Jupiters, indicating higher metallicity may not be necessary to form super Jupiters \citep{Swastik_2021}. 
  
The enhanced abundance of volatile elements as compared to refractory elements was first observed in the solar atmosphere  \citep{melendez}, this could be used as a possible signature of the solar system among solar twins \citep{ramirez2009,melendez2012}. However, high precision differential abundances of solar analogs and stellar twins in binary systems did not show a significant difference in the trend of stellar abundance and condensation temperatures among planet hosts and non-hosts  \citep{gonzalez_1,hernandez_2010,hernandez_2013,planet_li_eu}. In fact, \citet{adibekyan2014f} noticed a significant correlation of the stellar abundances versus condensation temperature slope with stellar age and Galactocentric distance among Sun-like stars, which could be a cause for the observed difference in the volatile and refractive element abundances. Stellar lithium abundance could be a sensitive indicator of planet pollution; however, the results were inconclusive, showing a large spread in Li even among stars of very similar stellar parameters  \citep{pollak,Israelian2009,gonzalez_1,Figueira_2014,gonzalez_2014,Li_Eu,Delgado_Mena_2014,Delgado_Mena_2015}.
Carbon is produced in massive stars similar to $\alpha$ elements at low metallicities, but low-mass AGB (Asymptotic Giant Branch) stars could also make carbon \citep{origin_carbon,Kobayashi_2020} at higher metallicities, and hence the C/O ratio can change with time.  \citet{bond_2010a, delgado_mena_2010} showed the importance of C/O ratio in the formation of carbide and silicates in the planet formation and determine the planet mineralogy \citep{co_nikku}. 

\citet{c_n_o_s} studied 91 planet hosts and 31 non-host solar-type dwarf stars using atomic carbon lines and found no significant difference in [C/Fe] for the planet host and the non-host stars.  \citet{delgado_mena_2010}  also found no difference between carbon abundance between giant planet hosts and non-host stars. \citet{ cno_ch_band} used the CH band at $4300 \AA$ for deriving the carbon abundance instead of the atomic lines at $5380.3 \AA$ and $5052.2 \AA$ to study the carbon abundance of HARPS FGK stars with 112 giant planet hosts and 639 stars without known planets. Furthermore, they found that [C/Fe] is not varying as a function of the planetary mass, indicating the absence of a significant contribution of carbon in the formation of planets.


In this paper, we present the occurrence rate analysis of carbon abundance based on a large number of Kepler-LAMOST (The Large Sky Area Multi-Object Fiber Spectroscopic Telescope) samples of main-sequence  FGK stars to understand the importance of carbon abundance in the context of planet formation process as well as GCE using CH G band at 4300$\AA$. The sample contains 825 confirmed planet host stars and 214 stars with planet candidates from the Kepler catalog, and 49215 stars without detected planets so far. 
\section{Data and target selection} \label{sec:data} 
\begin{figure}
\includegraphics[width=0.53\textwidth]{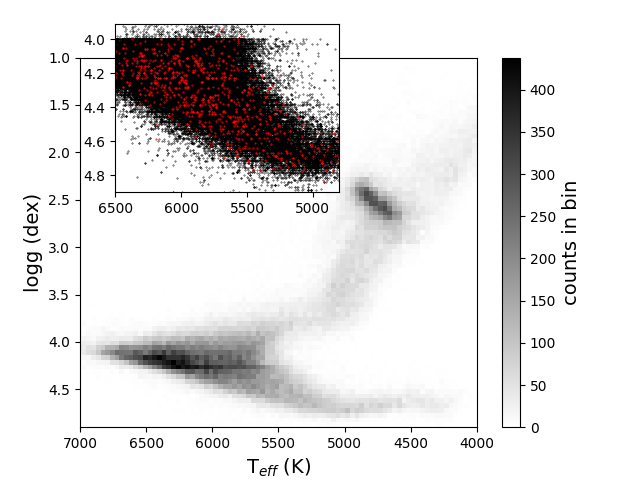}
\caption{The entire sample from the LAMOST-KEPLER field is shown as a density plot. And the selected samples within the restricted stellar parameters for the current study is shown as an inset. The black and red dots indicate the field and the planet host stars,  respectively.}
\label{hr1}
\end{figure}

\begin{figure}
\includegraphics[width=0.5\textwidth]{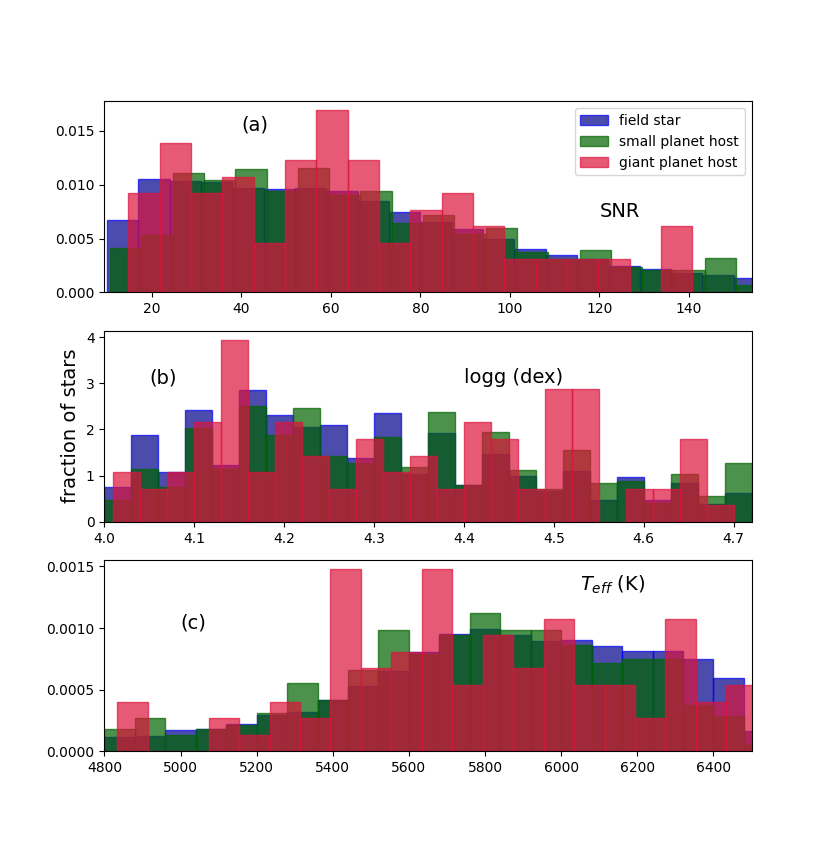}
\caption{ Distribution of SNR, $T_{eff}$ and \logg\ of the final sample (4800 $\leq$ $T_{eff}$ $\leq$ 6500 K and \logg\ $\geq$ 4.0) . The small planet host with planet radius R$_{p}\leq 4$R$_{\oplus}$, giant planet(R$_{p}>4$R$_{\oplus}$) host stars and field stars are in green, red and blue respectively.}
\label{teff_logg_snr}
\end{figure}

LAMOST is a wide field spectroscopic survey facility using a telescope with a 4m clear aperture and $5\deg$ field of view. The survey obtains 4000 spectra in a single exposure to a limiting magnitude of r=19 at the resolution R=1800 and simultaneous wavelength coverage of 370 - 900 nm \citep{Zhao2012}. We have used the LAMOST-Kepler project  \citet{Zong_2018} Public Data Release 4 (DR4) \footnote{ LAMOST DR4 complete data : \href{link:}{http://dr4.lamost.org/}} data for the current study. The observations were carried out between 2012 and 2017 and covered the entire Kepler field. A total of 227870 spectra belonging to 156390 stars were available in the database and out of which the spectroscopic parameters for 126172 stars were available from the LASP pipeline \citep{lamost_data_reduction}. The spectra and the corresponding stellar parameters (e.g.,  T$_{eff}$, log$\,g$, $[Fe/H]$ and radial velocity) were obtained from the LAMOST database. Additional parameters such as the mass and the radius of the planets are taken from the NASA Exoplanet archive \footnote{NASA exoplanet archive : \href{link:}{https://exoplanetarchive.ipac.caltech.edu}} \citep{exoplanet_archive}.
 We restricted the analysis to the main sequence stars ($4800 \leq T_{eff} \leq 6500$ K and \logg\ $\geq 4.0$ ), leading to a final sample of 49215 field stars and 1039 host stars with conformed exoplanets and potential candidates. Figure \ref{hr1} shows the parameter range of the final LAMOST-Kepler sample. Figure \ref{teff_logg_snr} shows the SNR (Signal to Noise Ratio), \logg\ and $T_{eff}$ histogram distribution of the final sample.

\begin{figure*}
\centering
\includegraphics[width=1.1\textwidth]{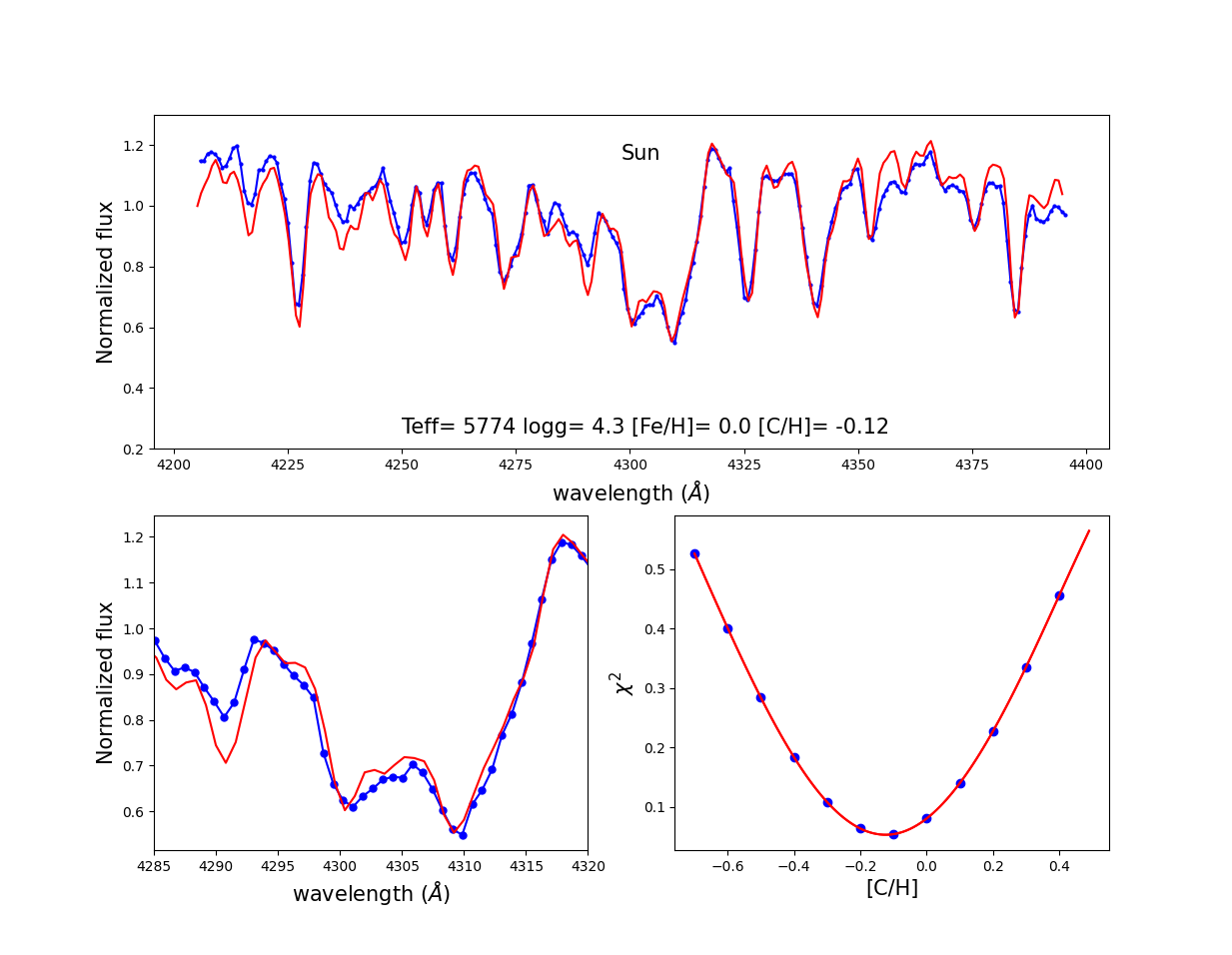}
\caption{Observed spectra of Sun (blue) and the synthetic spectra (red). Top panel shows the best fit with input parameters from LAMOST. Bottom-left panel shows an enlarged view of the CH G-band region. Bottom right panel shows the $\chi^{2}$ variation with  $[C/H]$ for obtaining the best fit $[C/H]$ value for the Sun with $T_{eff} = 5774$ K and SNR= 76.}
\label{spe1}
\end{figure*}

\begin{figure*}
\centering
\includegraphics[width=1.1\textwidth]{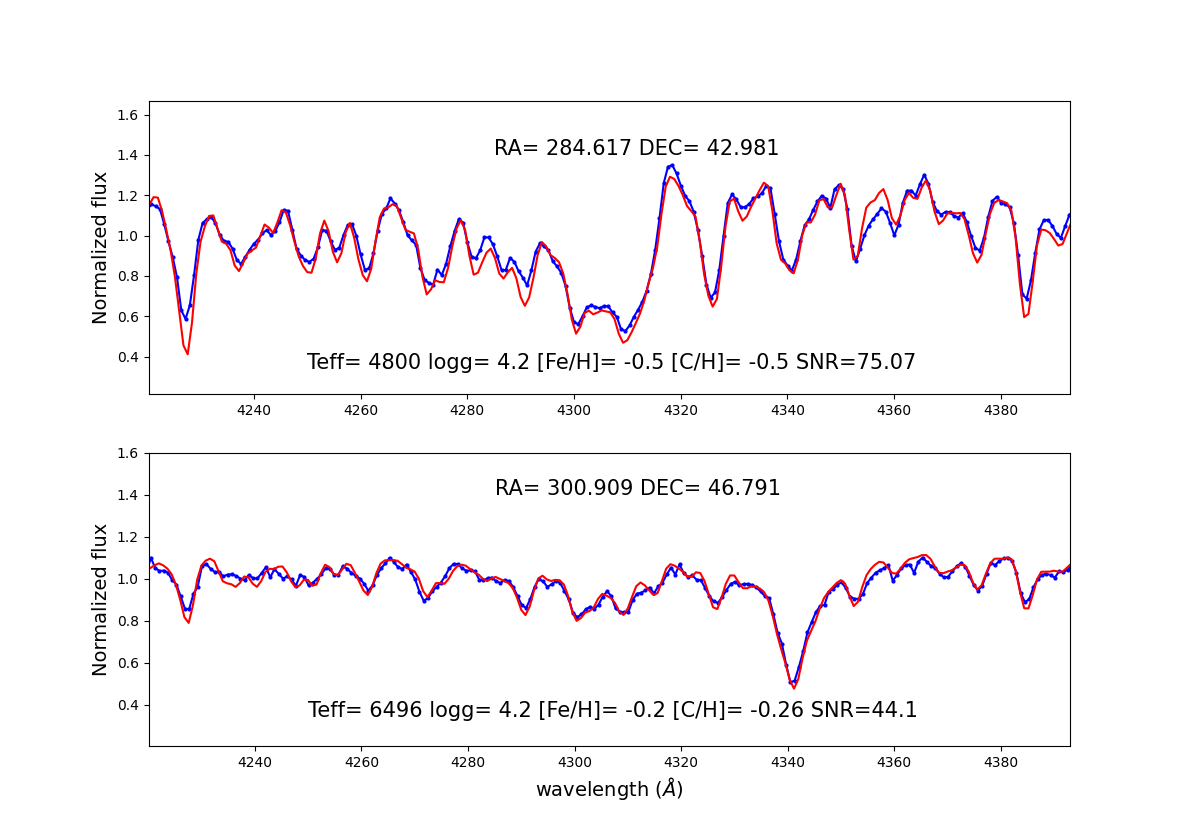}
\caption{ Observed spectra (blue) and synthetic spectra (red). Top panel shows the best fit for a star of $T_{eff} = 4800$ K and SNR= 75.07. Bottom-left panel shows for the best fit for a star with $T_{eff} = 6496$ K and SNR= 44.10.}
\label{spe2}
\end{figure*}

\section{Estimation of carbon abundances}
The methodology for estimating the carbon abundance uses a grid of synthetic spectra of varying carbon abundances across various stellar parameters and interpolates the model spectra to match the observed spectra. In this work we used Kurucz ATLAS9 NEWODF  \citep{kurucz_model} stellar atmospheric models by \citet{castelli2003} and Turbospectrum \citep{turbo_spectrum_1} spectrum synthesis code V19.1 \citep{plez2012} for generating the synthetic spectra . The atomic and molecular line lists are the same as that of \citet{lee2008} and \citet{carollo2012} with minor updates to the hyperfine structure and inclusions of isotopes for the heavy elements. The synthetic grid covers a wavelength range  4200-4400\AA, which covers the CH molecule of the G-band region, which is sensitive to carbon abundance. The synthetic spectra cover a range in effective temperatures between \teff $= 3500 - 7000$ K, with an increment of $250$ K and \logg\ range is between $0.0 - 5.0$ dex with an increment of 0.5 dex and $[Fe/H] = -1.0 - +0.5$ dex (with 0.5 dex increment). Carbon abundance was varied over this stellar parameter range at every $0.1$ dex step size. We used a python script for interpolation and $\chi^{2}$ minimization between the observed and the model spectra. Since the wavelength coverage of the grid is limited, stellar parameters from LAMOST were used, and only the carbon abundances are varied for estimating the best fit between observed and synthetic spectra. Figure \ref{spe1} shows an example of a best-fit spectrum. Solar scaled abundances are used for the stellar model atmospheres and synthetic spectra generation in the range $[Fe/H] = +0.0 - +0.5$ and for the metal-poor range, $[Fe/H] = -1.0 - -0.5$ dex an alpha enhanced abundances of $[\alpha/Fe] = 0.4$ dex was used. Solar abundances values are taken from \citet{grevesse2007} where $log(N(C)/N(H))+12=  8.39$ and $log(N(O)/N(H))+12=8.66$ were used. The synthetic spectra grid uses an oxygen abundance ($[O/H]$) that scales with the metallicity for the metal-rich models ($0.0 < [Fe/H] < 0.5$) and follows the alpha abundance in the metal-poor models ($[Fe/H] < 0.0$), as expected by the GCE. We checked the sensitivity of the derived carbon abundances to the assumed oxygen abundance and found it has less impact on the current sample, as the targets have $T_{eff}>$ 4800K and C/O $<$ 1.0. We also visually inspected the goodness of the spectral fit for the entire planet host stars, using plots similar to Figure \ref{spe1}. Figure \ref{spe2} represents the goodness of the fit at two extreme $T_{eff}$ regime.
\section{Carbon abundances} 
\label{sec:carbon_abu}
\begin{figure}
\includegraphics[width=0.5\textwidth]{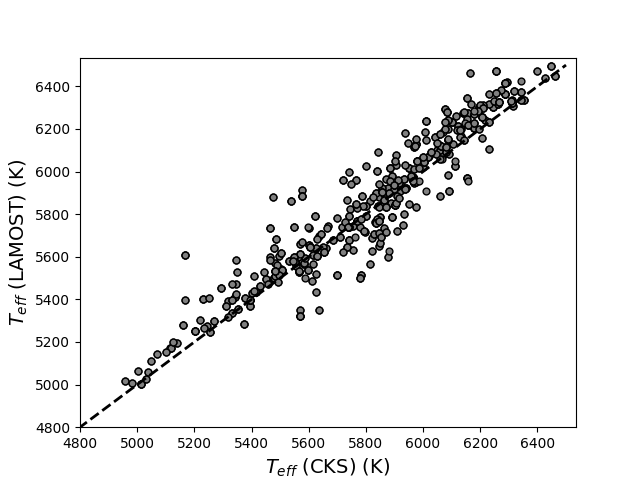}
\caption{ Comparison of the $T_{eff}$ values from CKS and the LAMOST. The black line is the 1:1 line.} 
\label{teff_teff}
\end{figure}

\begin{figure}
\includegraphics[width=0.45\textwidth]{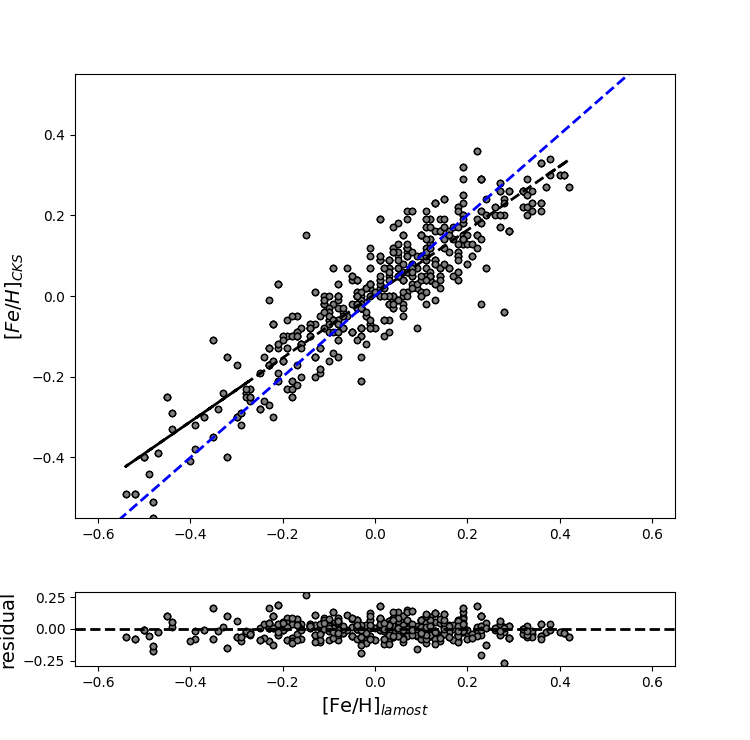}
\caption{Comparison of the $[Fe/H]$ values from CKS and the LAMOST pipeline. A linear fit is established between the CKS and the LAMOST $[Fe/H]$ values. The best fit coefficients are $ [Fe/H]_{new}=[Fe/H]_{lamost}*0.791+0.005$. The blue dashed line is the 1:1 line and black dashed line is the best fit line. } 
\label{cks-lamost_fit1}
\end{figure}

\begin{figure}
\includegraphics[width=0.5\textwidth]{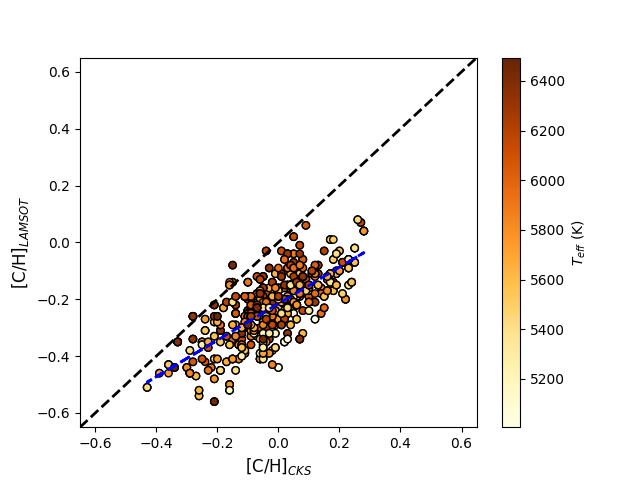}
\caption{ Comparison of the $[C/H]$ values from CKS and the LAMOST as a function of $T_{eff}$. The black dashed line is the 1:1 line and blue dashed line is the best fit line.} 
\label{cks-lamost_fit3}
\end{figure}

\begin{figure}
\includegraphics[width=0.5\textwidth]{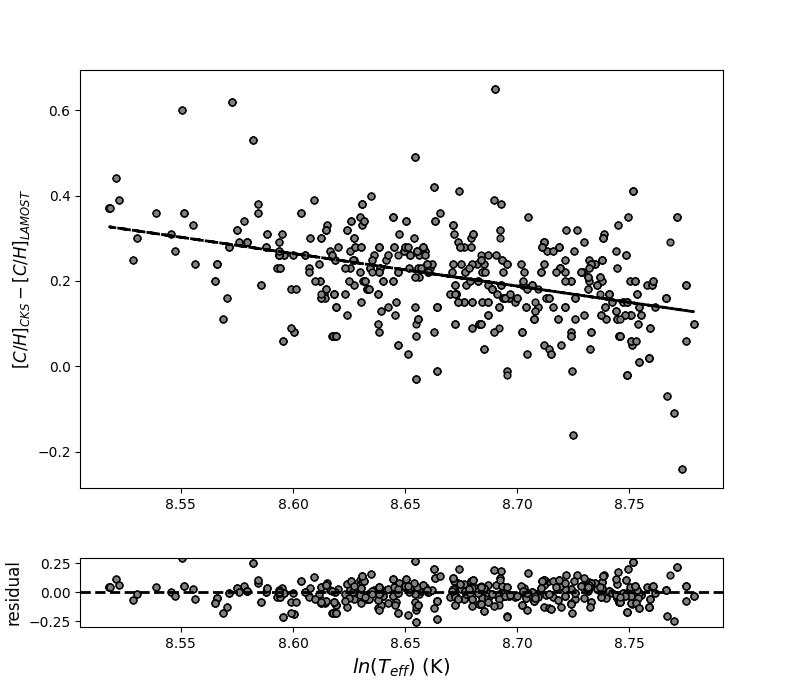}
\caption{ Comparison of $\delta [C/H]= [C/H]_{CKS}-[C/H]_{LAMOST}$ as a function of $T_{eff}$. A linear fit is established and the best fit coefficients are $\delta [C/H]=-0.762*\ln (T_{eff}(lamost))+6.820$ }.  
\label{cks-lamost_fit2}
\end{figure}

The carbon abundances derived in this work use low-resolution spectra fitting the strong CH feature. We corrected the LAMOST carbon abundances using common samples from the California Kepler Survey (CKS) \footnote{The CKS sample:  \href{DOI:}{https://doi.org/10.3847/1538-4365/aad501}}\citep{Brewer18}. We have compared the derived carbon abundances with previous studies and found that the trend in carbon abundances with respect to [Fe/H] is consistent with APOGEE (The Apache Point Galactic Evolution Experiment) \citep{apogee-kepler} and HARPS (High Accuracy Radial velocity Planet Searcher) \citep{delgado_mena_2010} data. We used  1025 common targets from CKS \citep{Brewer18} for deriving the corrections. As shown in Figure \ref{teff_teff}, the temperature scale between CKS and LAMOST common samples matches well after removing the 5$\sigma$ outliers. First we made corrections to the CKS and LAMOST [Fe/H] estimates (from the LAMOST catalog) , which is not significantly different ( Figure \ref{cks-lamost_fit1}). The LAMOST and CKS, [C/H] values show some dependency with effective temperature (Figure \ref{cks-lamost_fit3}).  So, in the next step, we derive corrections for [C/H] values as a function of  $T_{eff}$ (Figure \ref{cks-lamost_fit2}). We verified the correction for Sun using HARPS solar spectra. We have used Sun as star spectra from HARPS and convolved and re-binned to LAMOST resolution. We also added Gaussian noise to the data with an SNR=76, which is the mean SNR of the final sample. The stellar parameters we adopted for Sun are,  $T_{eff}=$5774 K, \logg\ = 4.3 dex and [Fe/H]=0.0. We found an offset of $[C/H]_{LAMOST}=-0.12 $ at solar temperature, which is consistent with the CKS corrections. The derived solar carbon abundance with CKS correction is $[C/H]_{LAMOST}=0.09 $  (Figure \ref{spe1}). In the following sections, we only use the CKS corrected LAMOST $[Fe/H]$ and $[C/H]$ values.  The CKS corrected [Fe/H], [C/H], along with the stellar parameters of the complete sample stars, are given in Table \ref{complete_sample}.

\begin{figure}
\includegraphics[width=0.45\textwidth]{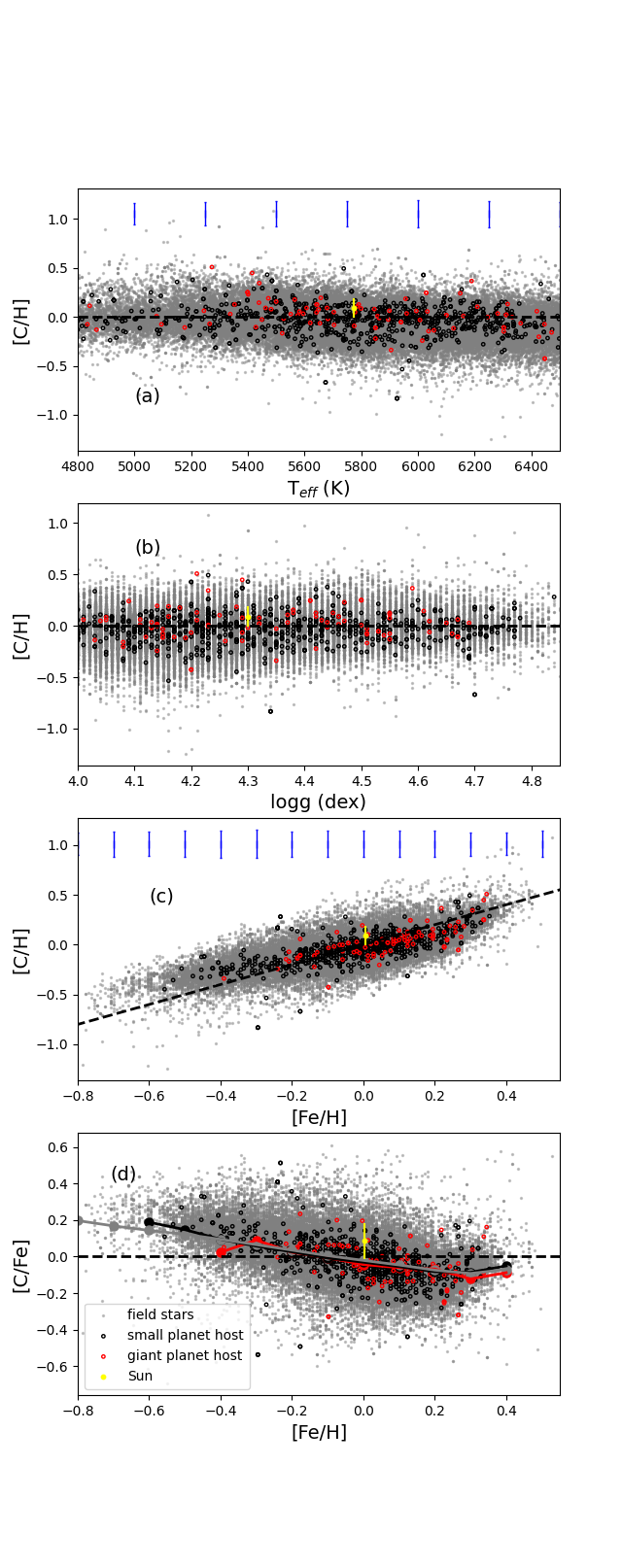}
\caption{Variation in $[C/H]$ as a function of T$_{eff}$ (a), log$\,g$ (b) and $[Fe/H]$ (c) after CKS correction. Giant planet host stars (red), Small planet host stars (black), field stars (grey) and Sun (yellow). Variation in $[C/H]$ error (shifted for visual purpose at different x values plotted in blue. In Figure \ref{teff_ch_logg} (c), the black dashed line show the 1:1 correlation. (d)  shows [C/Fe] as a function of [Fe/H] and the solid lines indicating the mean value of [C/Fe] in [Fe/H] bin of 0.2 dex for Giant planet host stars (red), Small planet host stars (black) and field stars (grey) .} 
\label{teff_ch_logg}
\end{figure}

We plotted the derived carbon abundances with respect to the stellar parameters to infer any systematic trends among them. Figure \ref{teff_ch_logg} shows no obvious correlation between the derived carbon abundances with  T$_{eff}$ and log$\,g$. Figure \ref{teff_ch_logg} (a) and (b) also shows no systematic difference in the T$_{eff}$ and log$\,g$ distribution between giant planet host stars, small planet host, and the field stars. The derived mean errors in the carbon abundances across different stellar parameters are also shown in the plots. The error in the carbon abundance is estimated from the $\chi^{2}$ difference for a fixed difference $\delta[C/H] = \pm 0.1$ dex in the carbon abundance around the minimum $\chi^{2}$. 
Figure \ref{teff_ch_logg} (c) represents $[C/H]$ as a function of $[Fe/H]$, that shows a positive trend  between $[Fe/H]$ and $[C/H]$ as expected due to the GCE effect. Both $[Fe/H]$ and $[C/H]$ increase linearly from the low metallicity close to the solar value and then flatten. This is the typical behavior of $\alpha$ elements that indicate carbon is primarily produced due to massive stars. Carbon may start to increase slightly at the very metal-rich end due to carbon production from the AGB stars; however, it is not very clear. Figure \ref{teff_ch_logg} (d) represents the trend of $[C/Fe]$ as a function of $[Fe/H]$, which also represents the GCE effect of carbon with respect to iron.  Both field stars and host stars follow a similar trend. From Figure \ref{teff_ch_logg} (d), the mean value of [C/Fe] as a function of [Fe/H] shows that, small planet host stars are preferentially found around higher [C/Fe] value in the metal-poor side ($[Fe/H]< -0.2$) compared to the field stars.

\section{Results} \label{sec:results}
We study the distribution of carbon abundance for planets of different radii and the occurrence rates with respect to the metallicity and carbon abundances. Using Galactic velocity dispersion, we infer the ages of the sample independent of the chemical abundances to understand the role of planet formation on the chemical composition. 
\subsection{Elemental abundance of the host stars as a function of planet population}
\begin{figure}
\begin{center}
    \includegraphics[width=0.5\textwidth]{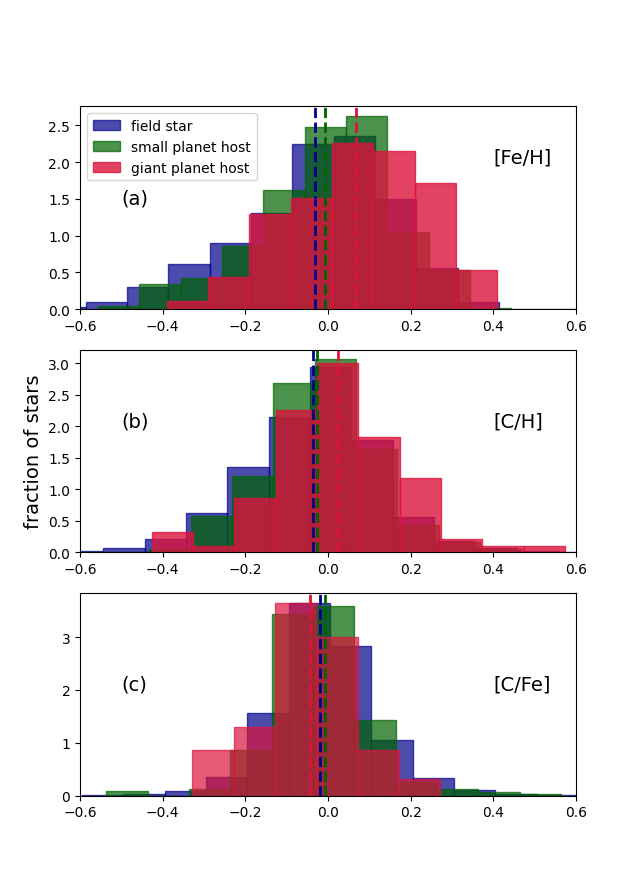}
    \caption{Distribution of carbon among small planet(R$_{p}\leq 4$R$_{\oplus}$) host, giant planet(R$_{p}>4$R$_{\oplus}$) host stars and field stars. Dashed vertical line represents the mean value of each distribution.} 
    \label{feh_distri}
\end{center}
\end{figure}

We examined the elemental abundance distribution of three distinct stellar populations: (i) host stars of the smaller planet (R$_{p} \leq 4R_{\oplus}$), (ii) host stars with giant planets (R$_{p} > 4R_{\oplus}$) and, (iii) Kepler field stars with no known planet detection. 
Distributions of $[Fe/H]$, $[C/H]$ and $[C/Fe]$ as a function of planetary radius is shown in Figure \ref{feh_distri}. We find that (a) giant planet hosts, on average, have a higher value of $[Fe/H]_{mean}$ as compared to the host stars of small planets and the field stars. This indicates that the giant planets are preferentially found around metal-rich host stars, which is similar to previous studies \citep{Mulders16, Petigura18, mayank}. And even the smaller planet hosts have a slightly higher $[Fe/H]_{mean}$ as compared to the field stars, perhaps indicating that for the formation of small planets $[Fe/H]$ could have some role \citep{small_radius_planet_feh1}.

Similarly in figure \ref{feh_distri} (b), the distribution of $[C/H]$ also follows similar trend as that of $[Fe/H]$. The giant planet host stars are carbon-rich compared to field stars and small planet host stars. The resulting $[C/H]$ trend is expected because $[C/H]$ increases with $[Fe/H]$ due to GCE. However, the difference between the $[C/H]$ distribution for small planet hosts and field stars is insignificant.
Figure \ref{feh_distri} (c) shows the distribution of $[C/Fe]$ for host stars of different planet radii. We find $[C/Fe]$ peaks at a higher value for the field stars compared to the planet hosts, which could be again due to the effect of GCE. Since most of the field stars are $[Fe/H]$ poor compared to the planet hosts, the $[C/Fe]$ at lower metallicities are expected to be higher, as most of the carbon in the Galaxy seems to have come from massive stars and hence the $[C/Fe]$ is high than solar values at lower metallicities \citep{origin_of_C_to_U}. Beyond solar metallicities, the rate of increase of iron is higher compared to carbon; hence the $[C/Fe]_{mean}$ value for the giant planet host star is low compared to stars hosting small planets and field stars. The results are shown in Table \ref{hist_result}. 

\begin{table}[htb]
\begin{center}
\small
\setlength\tabcolsep{0pt}\caption{Main results from the histogram distribution}\begin{tabular}{ccccrrrrr}
\hline\hline
category&$[Fe/H]_{mean}$&$[C/H]_{mean}$&$[C/Fe]_{mean}$ \\
\hline
field star&$-0.034\pm0.001$&$-0.036\pm0.001$&$-0.006\pm0.001$\\
small planet host&$-0.006\pm0.005$&$-0.025\pm0.005$&$-0.019\pm0.004$\\
giant planet host& $0.068\pm0.016$&$0.023\pm0.016$&$-0.044\pm0.012$\\ 
\hline \label{hist_result}
\end{tabular}
\end{center}
\end{table}

\subsection{Occurrence rate of planets as a function of host star abundance}

The analysis described in the previous sections does not take the completeness of the Kepler survey, the detector efficiency, or the probability of detecting a planet into account. The real trend can not be inferred from histograms. In order to derive the correlation between the host star elemental abundance and the planet radius that is free of selection effects and observational biases, we use the final $Kepler$ data release DR25 \footnote{ Kepler DR25  data : \href{doi:}{https://doi.org/10.3847/1538-4365/229/2/30}} catalog \citet{Mathur_2017} to compute the occurrence rate of exoplanets as a function of radius and host star $[Fe/H]$ and $[C/H]$. We updated the Kepler DR25 catalog with updated stellar and planetary radius based on Gaia DR2 from  \citet{Berger18}. Since the LAMOST metallicity and the derived carbon abundances are calibrated with respect to the CKS values, we combine CKS samples \citep{cks2,Brewer18} that has metallicities and carbon abundances. This also added additional samples for the occurrence rate estimation. To compute the occurrence rate as a function of planetary radii, we followed the prescription presented in \citet{Youdin11,Howard12,Burke15, Mulders16}.

\begin{figure}
  \centering
\includegraphics[height=5cm]{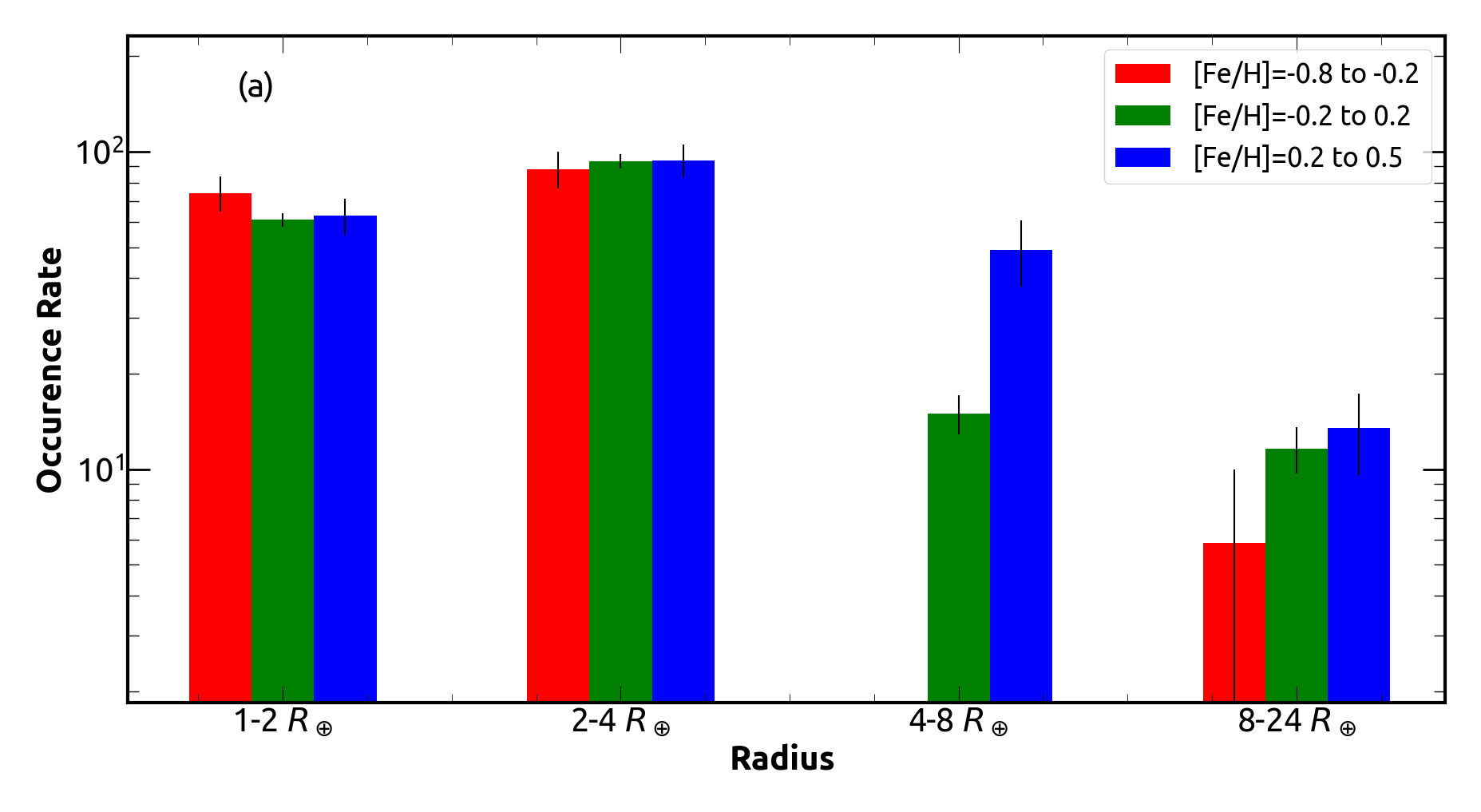}
  
\includegraphics[height=5cm]{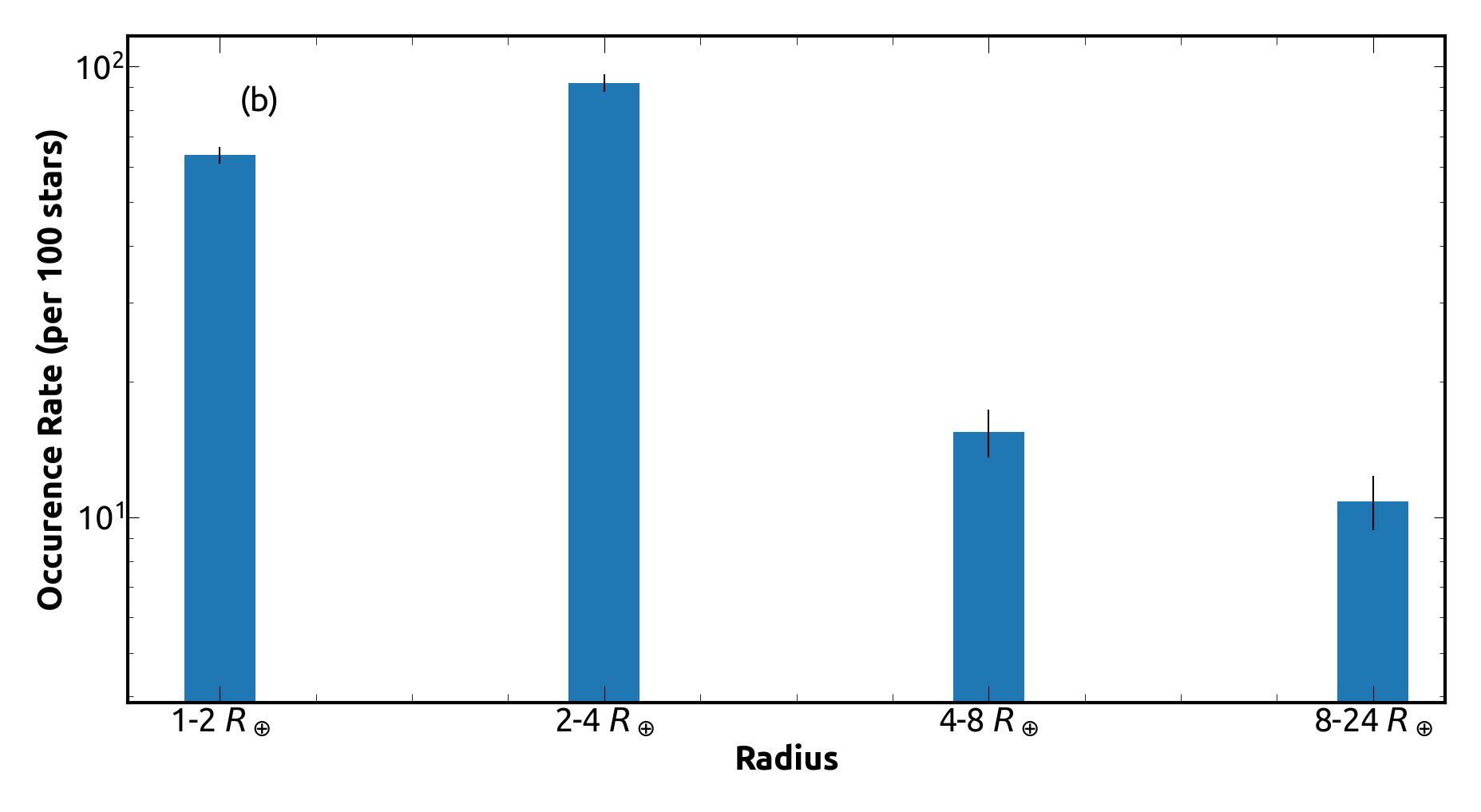} 
  
\includegraphics[height=5cm]{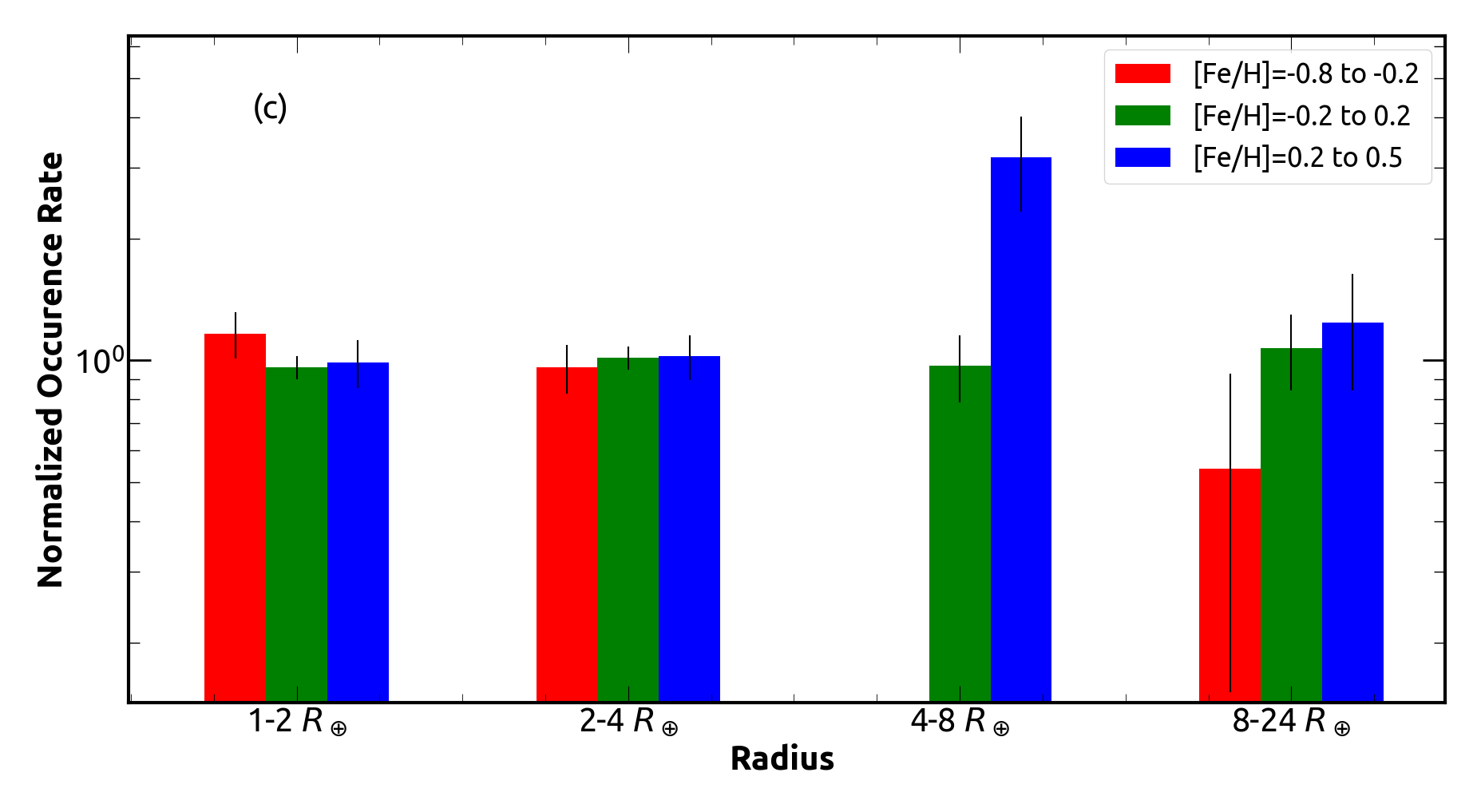}

  \caption{(a) Occurrence rate of exoplanets as a function of the planetary radius and the host star metallicity. (b) The total occurrence rate of the sample without subdividing it into different metallicity bins. (c) Normalized occurrence rate of the exoplanets as a function of the planetary radius and the host star metallicity. The error bars in these plots are the Poissonian errors based on the number of planets in each bin. }
\label{fig7}  
\end{figure}

Similar to \citet{mayank}, we have divided the sample into three $[Fe/H]$ bins; (i) sub-solar $[Fe/H]$ ($-0.8<[Fe/H]<-0.2$) , (ii) solar $[Fe/H]$ ($-0.2<[Fe/H]<0.2$) and (iii) super-solar $[Fe/H]$ ($0.2<[Fe/H]<0.8$). In Figure \ref{fig7} (a), the occurrence rate of the sample is shown as a function of planet radius and host star $[Fe/H]$. We also calculated the occurrence rate (for the exoplanet sample used in Figure \ref{fig7} (a) as a function of planet radius Figure \ref{fig7} (b).  The \ref{fig7} (a), is both a function of host star $[Fe/H]$ and radius. Similar to \citet{mayank}, we  normalized the occurrence rate in Figure \ref{fig7} (a) with the total occurrence rate as a function of radius Figure \ref{fig7} (b), to produce the normalized occurrence rate. The normalized occurrence rate Figure \ref{fig7} (c) is only a function of the host star $[Fe/H]$.  From Figure \ref{fig7} (a) and Figure \ref{fig7} (b) it can be seen that smaller planets $R_P \leq 4R_{\oplus}$ have similar occurrence rate for the three $[Fe/H]$ ranges, while giant planets $R_P > 4R_{\oplus}$ have a higher occurrence rate for the solar and super-solar $[Fe/H]$. This is consistent with the previous works in literature \citep[e.g.,][]{Mulders16, Petigura18, mayank}.   
 
\begin{figure}
  \centering
\includegraphics[height=5cm]{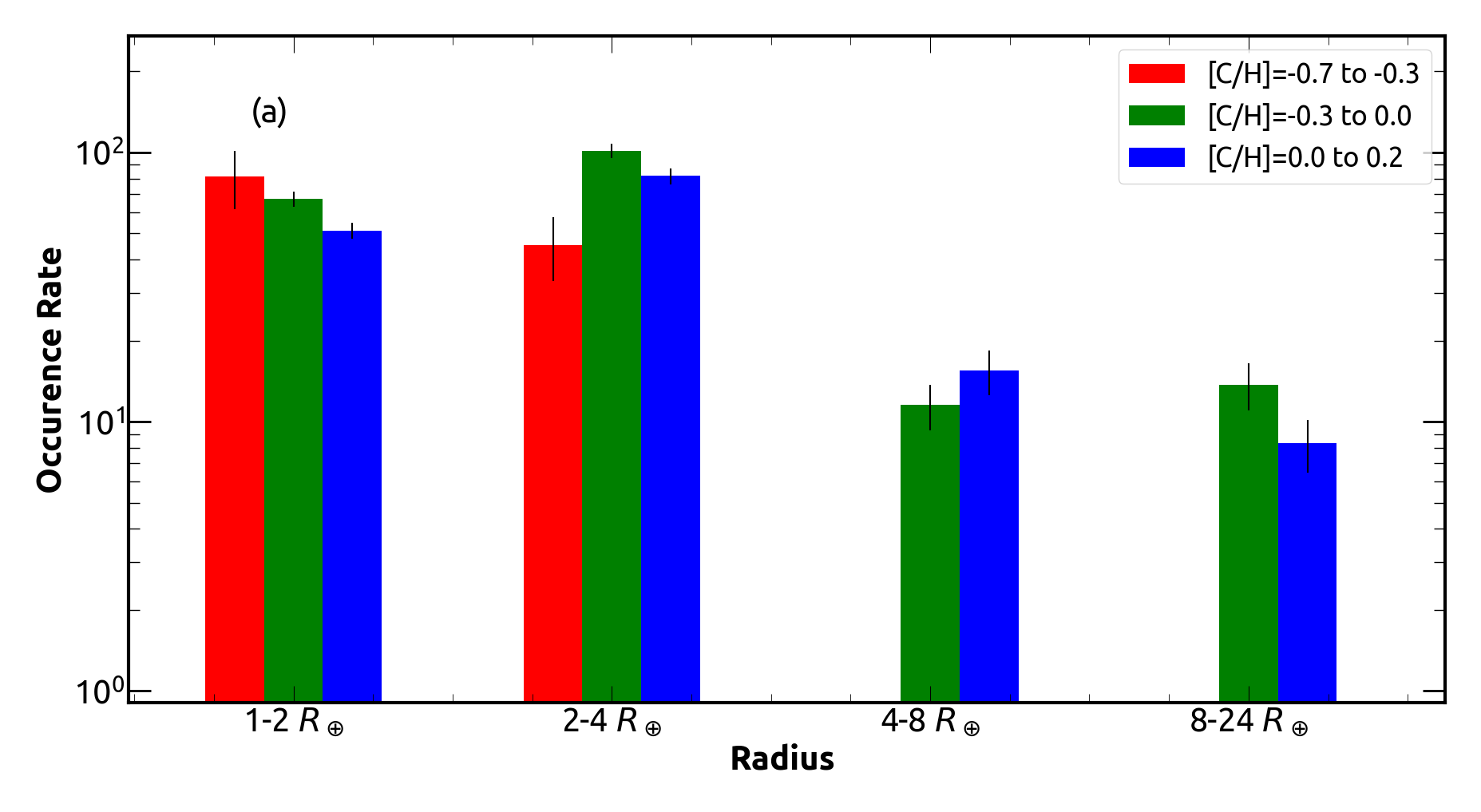}

\includegraphics[height=5cm]{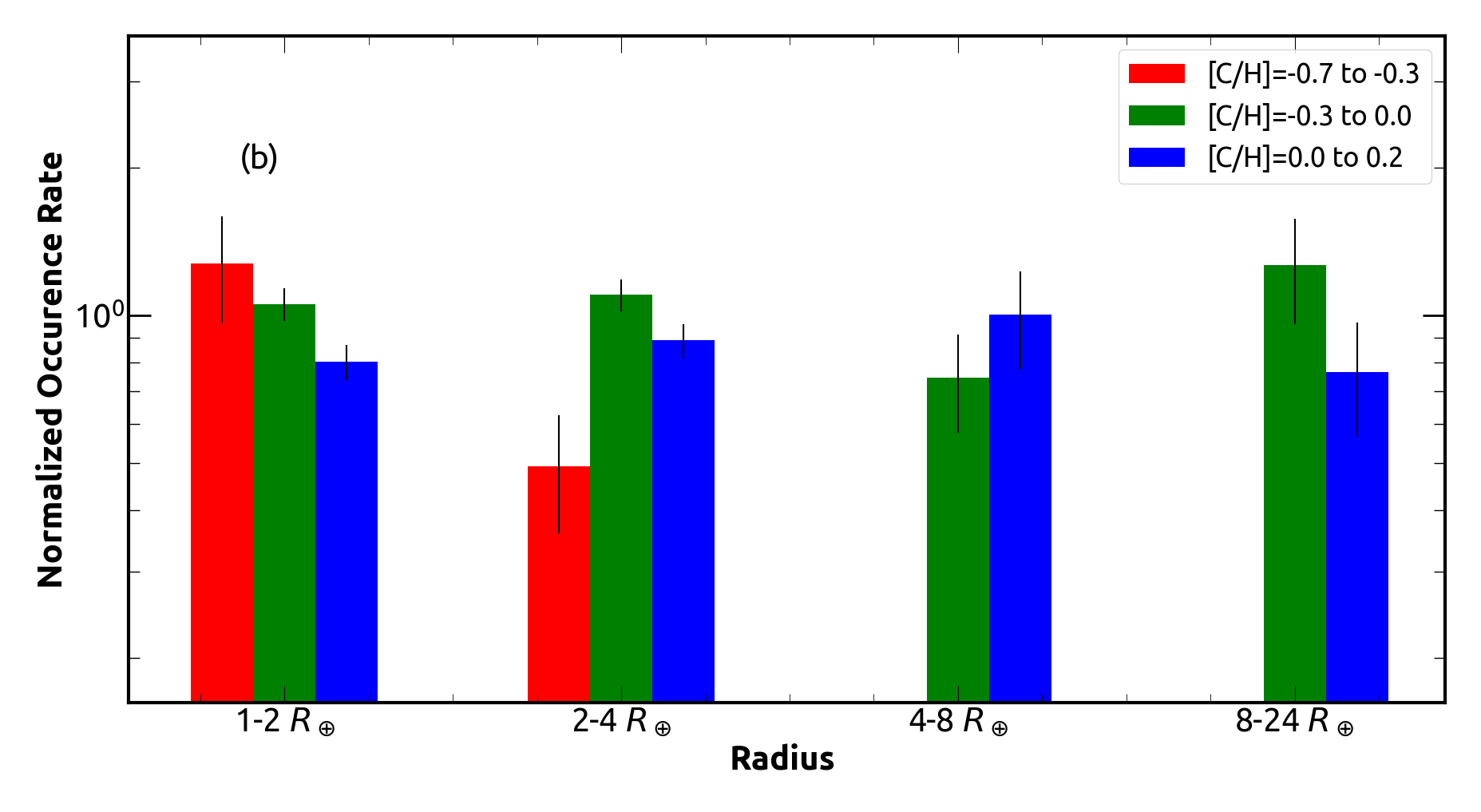}

  \par

  \caption{(a) Occurrence rate of exoplanets as a function of the planetary radius and the host star $[C/H]$. (b) Normalized occurrence rate of the exoplanets as a function of the planetary radius and the host star $[C/H]$. }
  \label{fig8}
\end{figure}

\begin{figure}
  \centering
\includegraphics[height=5cm]{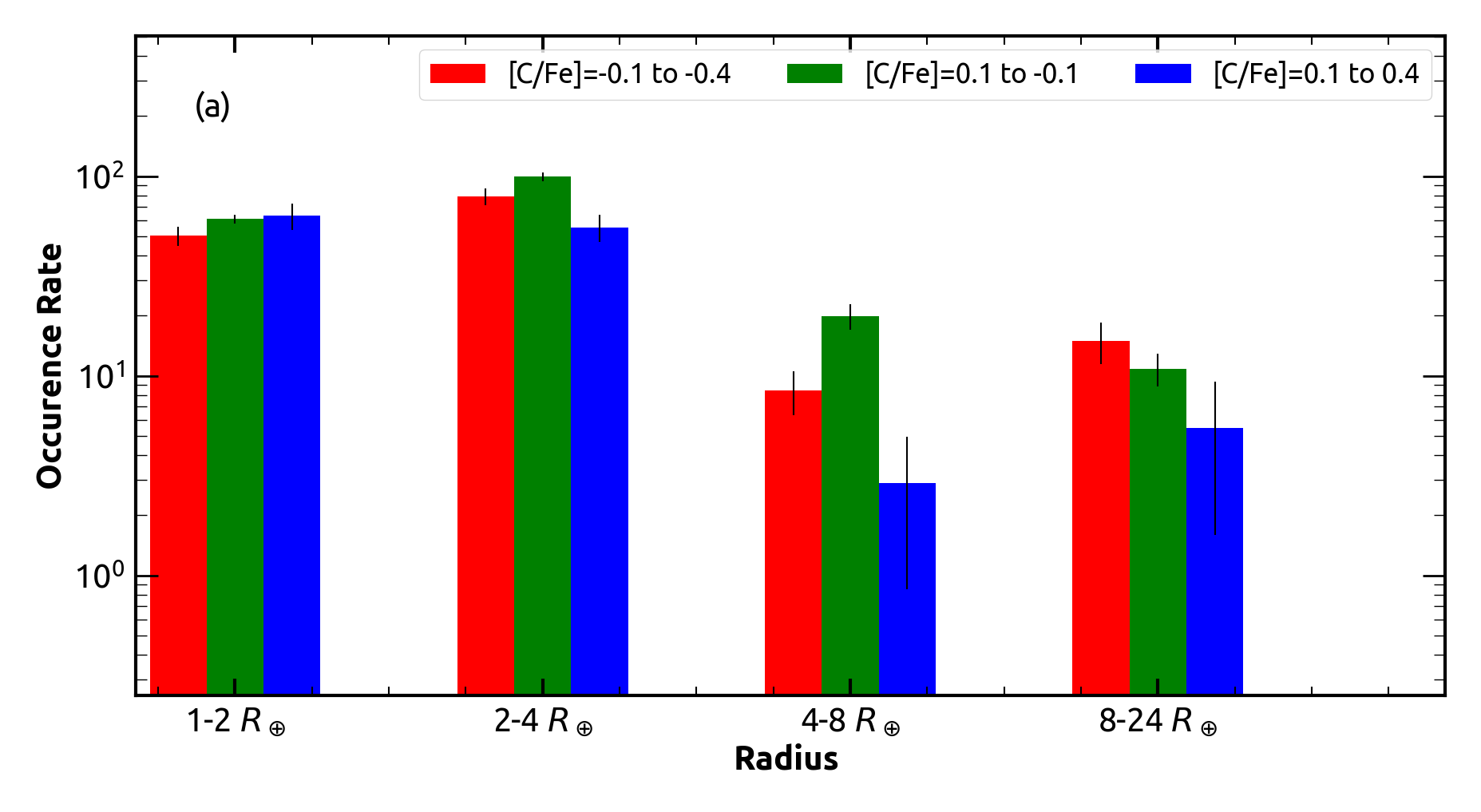}

\includegraphics[height=5cm]{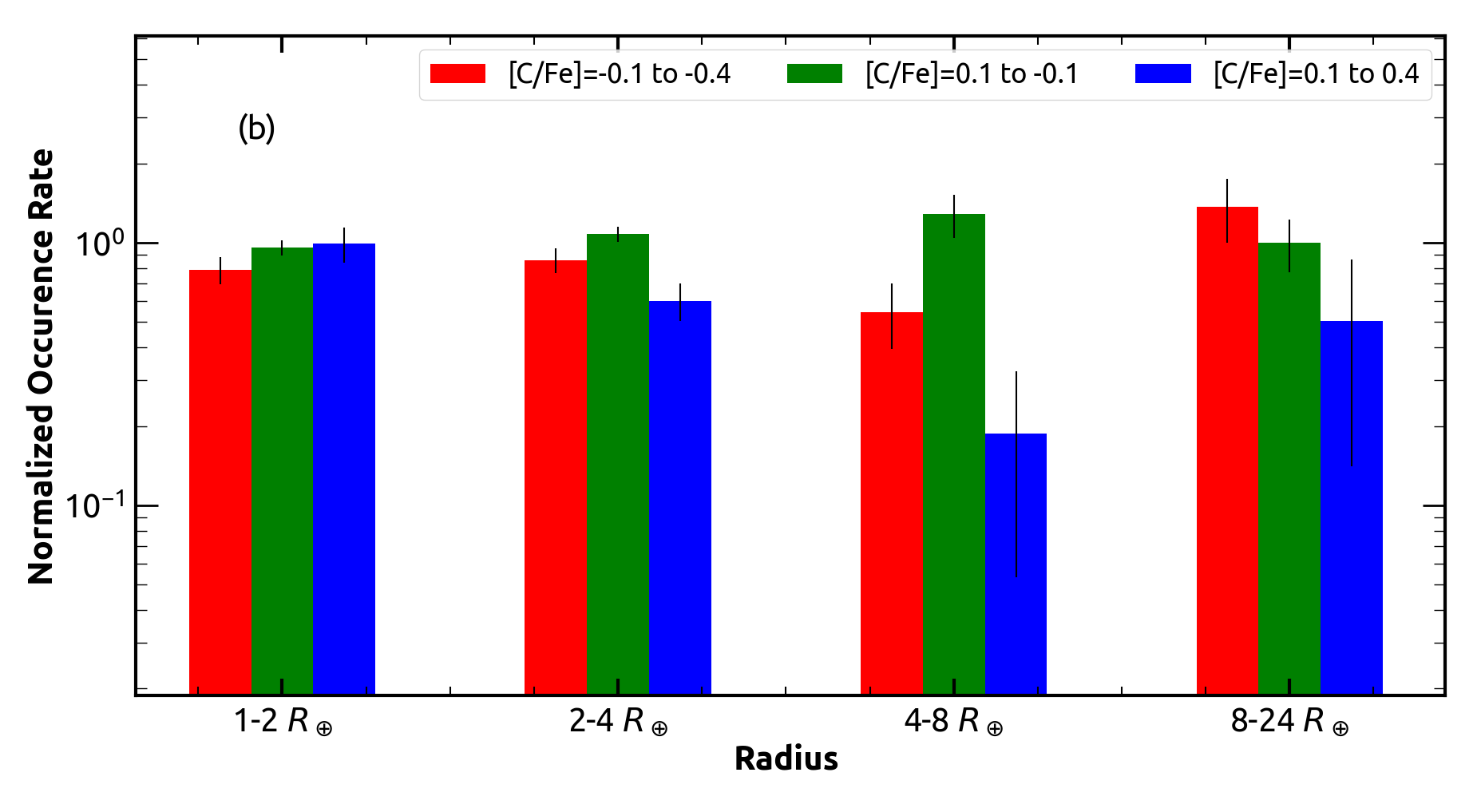}

  \par

  \caption{(a) Occurrence rate of exoplanets as a function of the planetary radius and the host star $[C/Fe]$. (b) Normalized occurrence rate of exoplanets as a function of the planetary radius and the host star $[C/Fe]$. }
  \label{fig9}
\end{figure}

To compute the occurrence rate of planets as a function of $[C/H]$, we divided the sample into three $[C/H]$ bins. Since we found that the [C/H] is a strong function of [Fe/H] (see Figure \ref{fig9} (c)) we converted the [Fe/H] bins to [C/H] bins. Based on equation 
\begin{equation}
    [C/H]=0.657*[Fe/H]-0.165
\end{equation}
we define the bins as (i) sub-solar $[C/H]$ ($-0.7<[C/H]<-0.3$) , (ii) solar $[C/H]$ ($-0.3<[C/H]<0.0$) and (iii) super-solar $[C/H]$ ($0.0<[C/H]<0.2$). In Figure \ref{fig8} (a), the occurrence rate as a function of host star carbon abundance and planetary radius is shown. Similar to Figure \ref{fig7} (a), Figure \ref{fig8} (a), is a strong function of both planetary radius and $[C/H]$. In Figure \ref{fig8} (b), the normalized occurrence rate of planets (using Figure \ref{fig7} (b)) as a function of $[C/H]$ is shown. From Figure \ref{fig8}, we find that similar to Figure \ref{fig7}, the occurrence rate of giant planets is higher for stars with solar and super-solar [C/H]. 

We further analyzed the occurrence rate of planets as a function of $[C/Fe]$. We divide the sample again into three bins (i) $[C/Fe]$ between -0.4 to -0.1, (ii)   $[C/Fe]$ between -0.1 to 0.1, and (iii) $[C/Fe]$ between 0.1 to 0.4. In Figure \ref{fig9} (a), the occurrence rate as a function of host star $[C/Fe]$ and planetary radius is shown. We found that the occurrence rate for smaller planets  ($R_P \leq 4R_{\oplus}$) is similar in all the three $[C/Fe]$ bins, while the occurrence rate of the giant planets ($R_P> 4R_{\oplus}$) is much higher for $[C/Fe] < 0.1$. 
This might indicate that volatile elements such as carbon do not play a significant role in the formation of giant planets.

\subsection{Galactic space velocity dispersion} \label{sec:velocity_dis}
The increase in (normalized) occurrence rate as a function of $[C/H]$ indicates that carbon enhancement is a necessary step in the Galactic context of planet formation, though it might not play a strong role as that of $[Fe/H]$ in determining the size/radius of the planet. To further understand the planet population in the Galactic context, we need to understand the dependence and evolution of planetary properties and host star properties as a function of the Galactic age. In Narang et al. (under review), we have established that the critical threshold of $[Fe/H]$ in ISM  that was necessary to form Jupiter-like planets was only achieved in the last 5-6 Gyrs indicating that the Jupiters only started forming in the last 5-6 Gyrs. Since the $[C/Fe]$ values are expected to change over the timescale of the Galactic thin disk, we further investigated if probing the Galactic evolution of the $[C/H]$ and/or the $[C/Fe]$ might provide us with clues about the Galactic evolution of planetary systems. Similar to \citet{Binney00,Manoj05,Hamer19} and Narang et al. (under review), we used the dispersion in the peculiar velocity of the stars as a proxy for the age of the stars in the Kepler field. We  estimated the velocity dispersion (a proxy for the age)  as function of $[Fe/H]$, $[C/H]$ and $[C/Fe]$. To compute the velocity dispersion, we first calculated the Galactic space velocity in terms of the U, V, $\&$ W space components following \citet{Johnson87,Ujjal}. The total velocity dispersion ($\sigma_{tot}$) for a particular  ensemble of stars is then given as the quadratic sum of the individual components of the velocity dispersion in that given ensemble such that

\begin{equation}
   \sigma_{tot} = \sqrt{\sigma_U^2+\sigma_V^2+\sigma_W^2}
\end{equation}

where $\sigma_U$, $\sigma_V$, and $\sigma_W$ are the velocity dispersion of the $U$, $V$, and $W$ components given in the same manner as: 

\begin{equation}
   \sigma_U^2 = \frac{1}{N}\Sigma_{0=i}^N (U_i-\Bar{U})^2
\end{equation}
here N is the number of stars. 

Furthermore velocity dispersion can then be converted to an average age of the stars following the formalism from  Narang et al., (under review):

 \begin{equation}
    \sigma_{tot}(\tau) = A \times \tau^{\beta} 
    \label{eq11}
\end{equation}
{where $\tau$ is the average age of the host stars in a bin, A is a constant and is equal to $21.5 \, km\, s^{-1} \, Gyr^{-0.53}$ and $\beta =0.53$.  }  

\begin{figure*}
  \centering
\includegraphics[width=1\linewidth]{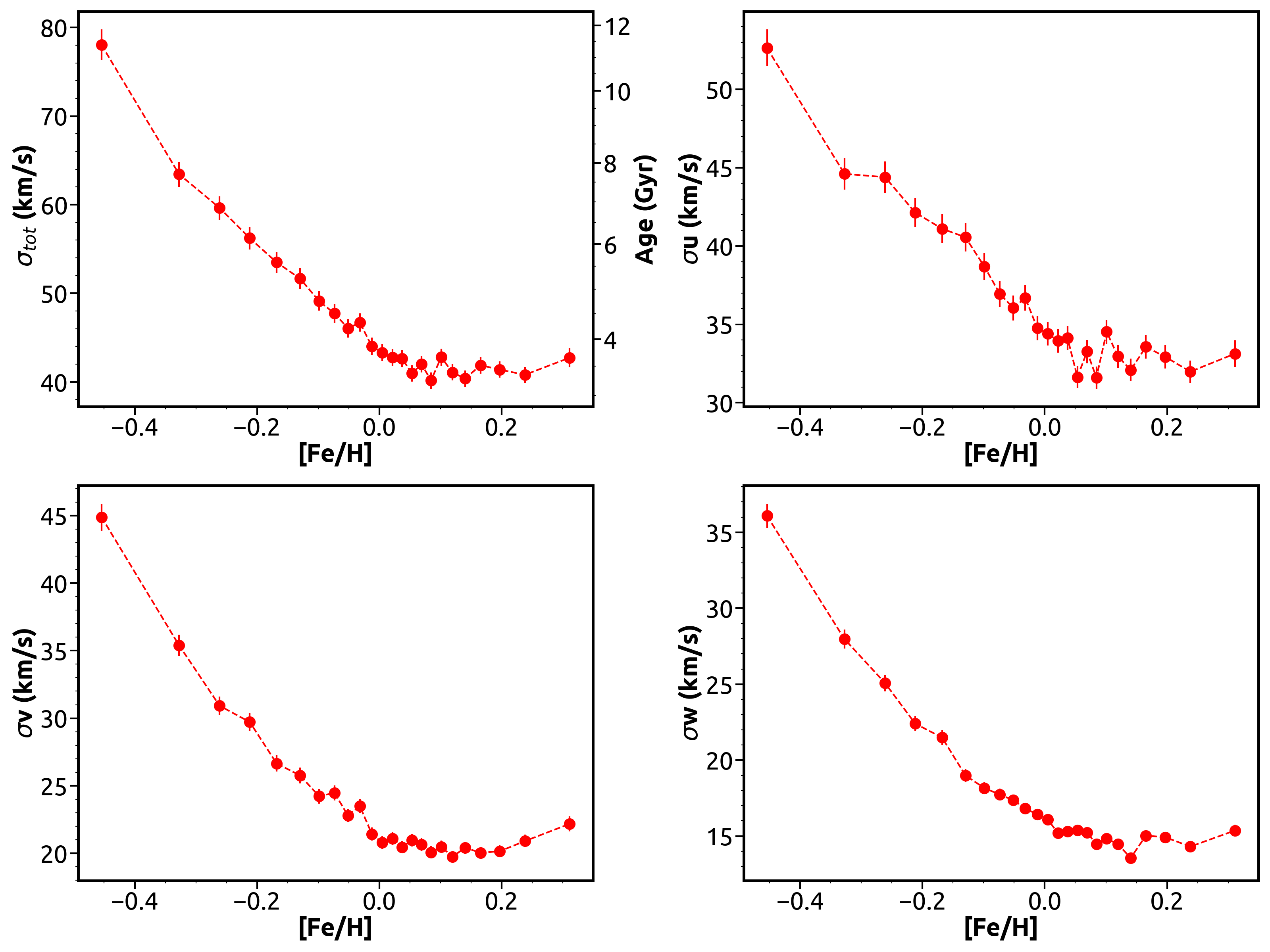}

  \caption{The total velocity dispersion and the velocity dispersion of the individual components ( $\sigma_{U}$,  $\sigma_{V}$ and  $\sigma_{W}$) as a function of $[Fe/H]$. We have used a running average with a bin size of 1000 and a step size of 200.  The error bar on the velocity dispersion are computed following \citet{Binney00}. On the right hand y-axis are the corresponding ages computed using Equation \ref{eq11}.  The same axis and error bar as well as binning scheme is followed for all subsequent figures. }
    \label{fig10}
\end{figure*}

In Figure \ref{fig10}, we show the velocity dispersion of the Kepler field as a function of $[Fe/H]$. As the average field $[Fe/H]$ increases, the total velocity dispersion $\sigma_{tot}$ decreases. This indicates that $[Fe/H]$ rich stars ($[Fe/H] > -0.2$) are younger. Similar behavior is seen for the $\sigma_{U}$,  $\sigma_{V}$, and  $\sigma_{W}$ as well. Using equation \ref{eq11}, we can further convert $\sigma_{tot}$ to the average age of the stars. We find that $[Fe/H]$ rich stars ($[Fe/H] > -0.2$) have an average age between $\sim$4-6 Gyrs. Further from Figure \ref{fig7}, we find that most giant planets are around $[Fe/H]$ rich stars. Hence from Figure \ref{fig7} and Figure \ref{fig10} we conclude that most of the giant planets ($R_P>$  4 $R_\oplus$) in the Kepler field are of an average age between $\sim$4-6 Gyrs, while smaller planets have a much larger spread in host stars $[Fe/H]$ and hence even in the age. 

\begin{figure*}
  \centering
\includegraphics[width=1\linewidth]{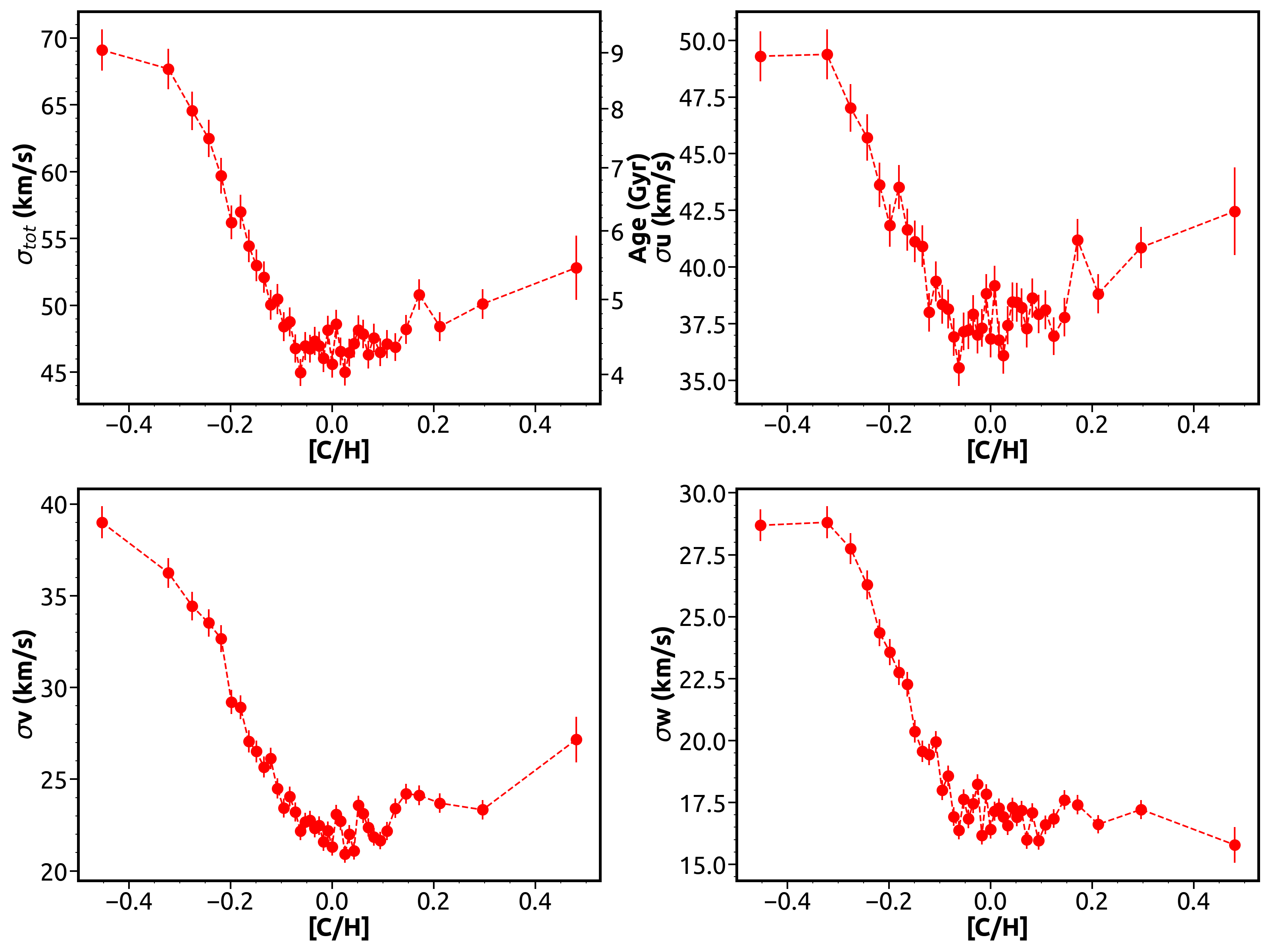}
  \caption{The total velocity dispersion and the velocity dispersion of the individual components ( $\sigma_{U}$,  $\sigma_{V}$ and  $\sigma_{W}$) as a function of $[C/H]$. }
    \label{fig11}

\end{figure*}

\begin{figure*}
  \centering
  \includegraphics[width=1\linewidth]{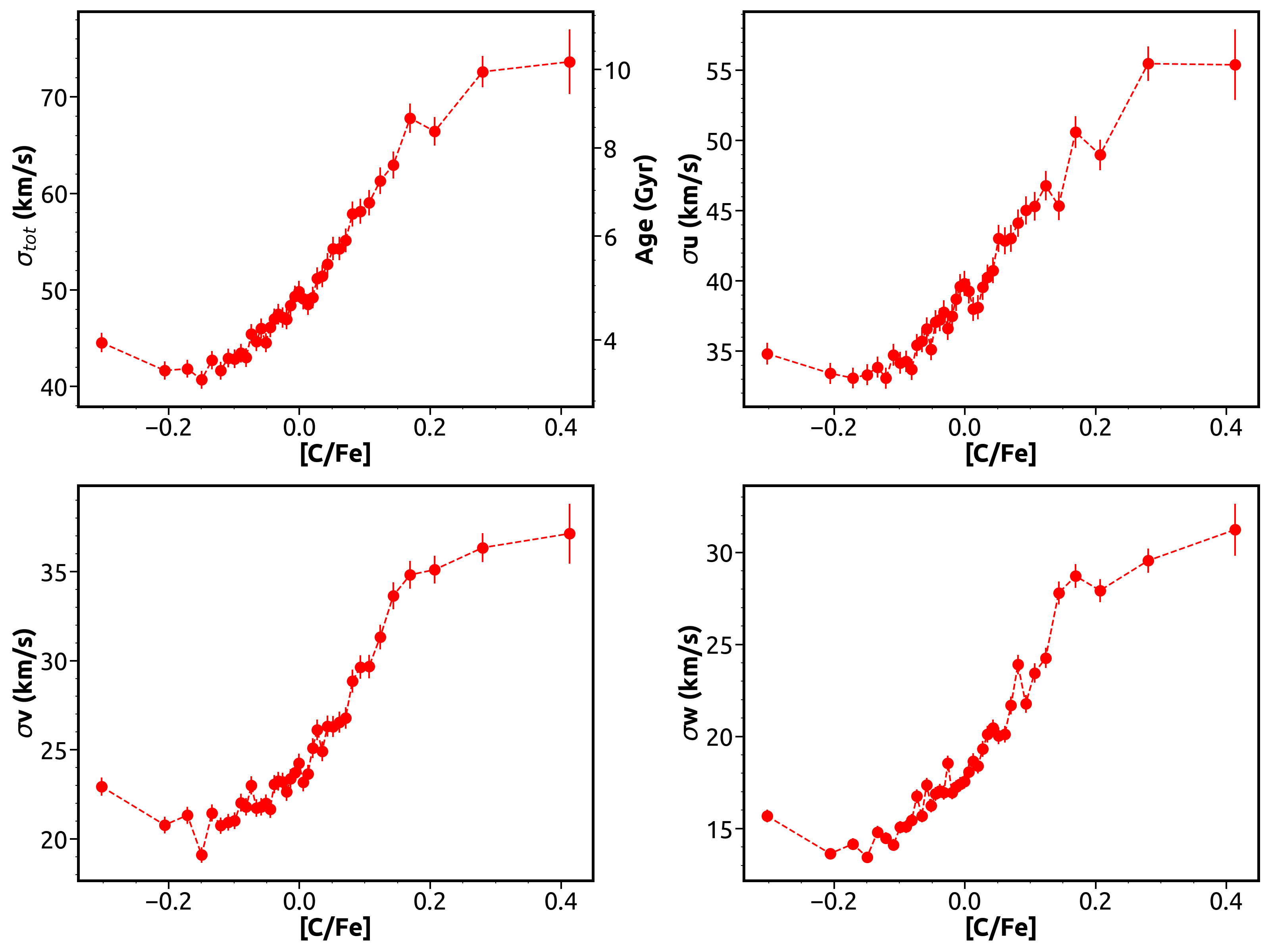}

  \caption{The total velocity dispersion and the velocity dispersion of the individual components ( $\sigma_{U}$,  $\sigma_{V}$ and  $\sigma_{W}$) as a function of $[C/Fe]$.}
    \label{fig12}
\end{figure*}

Similarly, by combining the results of velocity dispersion as a function of $[C/H]$ from Figure \ref{fig11} and the occurrence rate of planets in the Kepler field as a function of $[C/H]$, we find that the average age of host stars of giant planets $R_P>4R_\oplus$ is between 4-5 Gyrs. Similar age ranges are obtained based on $[C/Fe]$ as well (Figure \ref{fig12}).

\section{Discussion} \label{sec:discussion}
We have calculated the planet occurrence rate as a function of host star metallicity and carbon abundance. The distribution of $[Fe/H]$ and $[C/H]$ with respect to the planet radii show that planets with $R_{p}>4R_{\oplus}$ are preferentially found around stars with solar and super-solar metallicities.  At these preferred high metallicities, the GCE trend shows lower [C/Fe] ratios, and planet hosts also follow a similar trend as the field stars as shown in Figure \ref{teff_ch_logg}. With the current sample, we do not find a significant difference in the [Fe/H] versus [C/Fe] trend above solar metallicities between the field stars and planet hosts. We explored the difference in $[C/Fe]$ within a narrow bin in metallicity to remove the GCE trend; however, this has reduced the number of samples significantly. A simple mean gives a $[C/Fe]$ value of $-0.09$ for field stars and $-0.13$ for giant planet hosts for  $ [Fe/H] > 0.28$. However, at lower metallicities (where mostly low mass planet hosts are present), the planet hosts may have slightly higher [C/Fe] values than the field stars, which is similar to what is observed in alpha elements \citep{Adibekyan2012a}.
Hence, there may be a general preference for planet hosts to have higher abundances of metals. 
Since, planet hosts and field stars follow the GCE trends in elemental abundance, it is difficult to test the preference of a higher [C/Fe] among planet hosts at solar and super solar metallicities. Stellar population with different abundance ratios with overlapping metallicity, similar to thick and thin disk, is not present at higher metallicity. Giant planet frequency at a different Galactic distance from future microlensing surveys can cover a range of stellar metallicities and possibly with different abundance ratios.
\section{Conclusion}
We have used LAMOST-Kepler data of main-sequence dwarfs to derive the carbon abundance and compared the planet hosts and the field stars. We constrained the sample to the main sequence dwarf stars to avoid effects due to stellar evolution. The distribution of carbon and iron with the planet radii and the occurrence rate analysis showed that the giant planet hosts are metal-rich and carbon-rich compared to the field stars and the stars with smaller planets. However, at super-solar metallicities, the $[C/Fe]$ values are lower than the solar ratio. At the metal-rich end, iron increases at a faster rate compared to carbon, which may be crucial for increasing the abundance of the refractory elements. Based on the Galactic space velocity dispersion, we found that the Jupiter host stars are younger, only about 4-5 Gyrs old. From the detailed occurrence rate analysis, we found that carbon may not be a significant contributor to the mineralogy of planet formation as compared to iron.

\section{Acknowledgement}
We used LAMOST archival data and the NASA exoplanet archive for this study. Guoshoujing Telescope (the Large Sky Area Multi-Object Fiber Spectroscopic Telescope LAMOST) is a National Major Scientific Project built by the Chinese Academy of Sciences. Funding for the project has been provided by the National Development and Reform Commission. LAMOST is operated and managed by the National Astronomical Observatories, the Chinese Academy of Sciences. We thank and acknowledge the immense contributions of Castelli in generating an extensive grid of stellar atmospheric models that this work used.
\section{Appendix}

\begin{table*}[h]
  \centering
  \begin{tabular}{@{}ccccccc@{}}
  \hline 
    RA(degree)  & DEC(degree)    & $T_{eff}$ & \logg\ & [Fe/H]&[C/H]\\
    \hline
  12.9681  & 10.1142 &  5843.0 &  4.09&   -0.5487& -0.2814    \\
  13.1028 & 8.7679  &  6058.0  & 4.34  & 0.0207   &-0.0390   \\
  13.1220 & 9.1860&    5044.0  & 4.57& -0.4775&-0.2193   \\
  13.1915 &  9.5228   & 5562.0 &  4.25  &      0.0681 & 0.0960     \\
  13.2336 &  8.7201 & 5401.0& 4.57& -0.4854& -0.1415  \\
  13.2383 & 10.6629 & 5024.0 & 4.46& -0.0979& 0.2636       \\
  13.2399  &  9.5898 &   5444.0 &  4.44&   0.0444&-0.0075   \\
  13.2497  & 9.5943&    5803.0  & 4.04  &-0.5329& -0.3562\\
  13.2887 &  9.7260 &   6266.0& 4.11& -0.1928 & -0.1047     \\
  13.3265 &  9.8816 &   5361.0  & 4.42 &  -0.4696 & -0.2758    \\
  13.3561  & 11.1456 &  5858.0  & 4.44 &  0.0207& -0.0334   \\
  13.3963 &  9.6087  &  6201.0 &  4.03&  -0.0662&  -0.0568   \\
  13.4179&   9.7623  &  6465.0 &  4.15 &  -0.3668 &-0.3385    \\
  13.4460 &  9.7859 &   5720.0&   4.34 & -0.5566 &-0.5352 \\
    .......&.......&.......&.......&.......&.......\\
    .......&.......&.......&.......&.......&.......\\
    .......&.......&.......&.......&.......&.......\\
  303.2472 & 46.1495 &  5325.0   &4.42  & 0.4082  &    0.3392 \\      
  303.2586&  45.8985 &  6281.0  & 4.50 &-0.0662  &  0.1934 \\    
  303.2754 & 45.9066 &  5685.0 &  4.21 &  0.1472  &  0.0194 \\  
  303.2776 &  46.4944  &  5498.0 &  4.32 &  -0.3826 &-0.3450\\
  303.2888 & 45.2001 &  6111.0 &  4.17  & 0.0365  & -0.0256   \\
  303.3010 & 45.4655 &  5548.0 &  4.21& 0.1551 & 0.1180\\
  303.3096 & 45.9596 &  5933.0  & 4.36 &  0.0365  &  -0.0331  \\ 
  303.3115 & 45.8891 &  6059.0  & 4.26  & -0.0109 &   -0.0391\\
  303.3131 &  46.2382  & 5750.0 &  4.09 &  0.1156 &0.0007    \\
  303.3182 & 46.2866 &  6246.0 &  4.00&   -0.1058 & -0.0723    \\
  303.3222 & 45.8463 &  5833.0 &  4.00 &  -0.0662  &  -0.1701    \\
  303.3513 & 46.2210 &  6059.0  & 4.07 &  0.0760 & -0.0091   \\
  303.3543 & 46.0936 &  5697.0 &  4.42 & -0.0188 &  -0.0421  \\
  303.3627 & 45.4936 &  5605.0  & 4.05  & -0.3193 & -0.1597    \\
  303.3804 & 45.6495  &  5788.0&   4.40 & 0.1077 & -0.1142 \\
  303.3810 & 45.7701 & 6286.0 &  4.23   &-0.0188 &  -0.2871    \\
  303.4039 & 45.6290 &  5910.0 &  4.25   & -0.0109  &  0.1798\\
    \hline
  \end{tabular} 
  \caption{sample data \footnote{Complete data is available  }}
  \label{complete_sample}
\end{table*}

\bibliographystyle{aasjournal}
\bibliography{c_h_main}
\end{document}